\pdfoutput=1
\documentclass[reprint, amsmath, amssymb, aps]{revtex4-2}
\usepackage{color}
\usepackage{textcomp}
\usepackage{hyperref}
\hypersetup{colorlinks=false, hypertexnames=false}
\usepackage{graphicx}
\usepackage{dcolumn}
\usepackage{bm}

\begin{document}

\preprint{APS/123-QED}

\title{Mid-infrared spectroscopy with a broadly-tunable thin-film~lithium~niobate~optical~parametric~oscillator}

\author{Alexander Y. Hwang\textsuperscript{1}}%
\author{Hubert S. Stokowski\textsuperscript{1}}%
\author{Taewon Park\textsuperscript{1}}%
\author{Marc Jankowski\textsuperscript{2}}%
\author{Timothy P. McKenna\textsuperscript{2}}%
\author{Carsten Langrock\textsuperscript{1}}%
\author{Jatadhari Mishra\textsuperscript{1}}%
\author{Vahid Ansari\textsuperscript{1}}
\author{Martin M. Fejer\textsuperscript{1}}%
\author{Amir H. Safavi-Naeini\textsuperscript{1}}%

\affiliation{%
 \textsuperscript{1}E.L. Ginzton Laboratory, Stanford University, Stanford, CA, 94305, USA
}
\affiliation{%
 \textsuperscript{2}NTT Research, Inc., Physics \& Informatics Laboratories, Sunnyvale, CA, 94085
}

\begin{abstract}
Mid-infrared spectroscopy, an important and widespread technique for sensing molecules, has encountered barriers stemming from sources either limited in tuning range or excessively bulky for practical field use. We present a compact, efficient, and broadly tunable optical parametric oscillator (OPO) device surmounting these challenges. Leveraging a dispersion-engineered singly-resonant OPO implemented in thin-film lithium niobate-on-sapphire, we achieve broad and controlled tuning over an octave, from $1.5\text{--}3.3$~µm by combining laser and temperature tuning. The device generates ${>}25$~mW of mid-infrared light at $3.2$~µm, offering a power conversion efficiency of $15\%$ ($45\%$ quantum efficiency). We demonstrate the tuning and performance of the device by successfully measuring the spectra of methane and ammonia, verifying our approach's relevance for gas sensing. Our device signifies an important advance in nonlinear photonics miniaturization and brings practical field applications of high-speed and broadband mid-infrared spectroscopy closer to reality.
\end{abstract}

\maketitle



\section{Introduction}

A fundamental technique for sensing is mid-infrared (MIR) spectroscopy, which exploits molecules' strong and distinct absorption responses in the $2\text{--}20$~µm spectral region. High-sensitivity and high-resolution MIR spectroscopy with coherent sources has rich applications, \textit{e.g.}, in gas \cite{ycas_Highcoherence_2018}, chemical reaction \cite{hinkov_Midinfrared_2022}, and biological \cite{amenabar_Structural_2013} sensing. Further advancing broadband, field-deployable MIR sources would enable a multitude of applications in areas such as rapid portable health monitoring and wide-coverage greenhouse gas detection.

However, currently-available sources still suffer from significant limitations. For instance, compact quantum- and interband- cascade lasers have dramatically improved their output power and efficiency, making them prominent sources for MIR spectroscopy \cite{vitiello_Quantum_2015}. However, material-defined gain bandwidths restrict tuning to hundreds of $\text{cm}^{-1}$ \cite{meng_Broadly_2015}, limiting potential multi-species detection. Meanwhile, optical parametric oscillator (OPO) sources allow efficient conversion of low-noise, wavelength-agile near-IR lasers over extremely broad tuning ranges (often thousands of $\text{cm}^{-1}$) \cite{vainio_Midinfrared_2016, breunig_Continuouswave_2011}. However, their conventional use of bulk optics creates large footprints, high threshold powers, high cost, and demanding stabilization requirements. These factors limit widespread field applications of OPOs, despite many laboratory spectroscopic studies \cite{vainio_Midinfrared_2016}.

\begin{figure*}[t]
  \begin{center}
      \includegraphics[width=\textwidth]{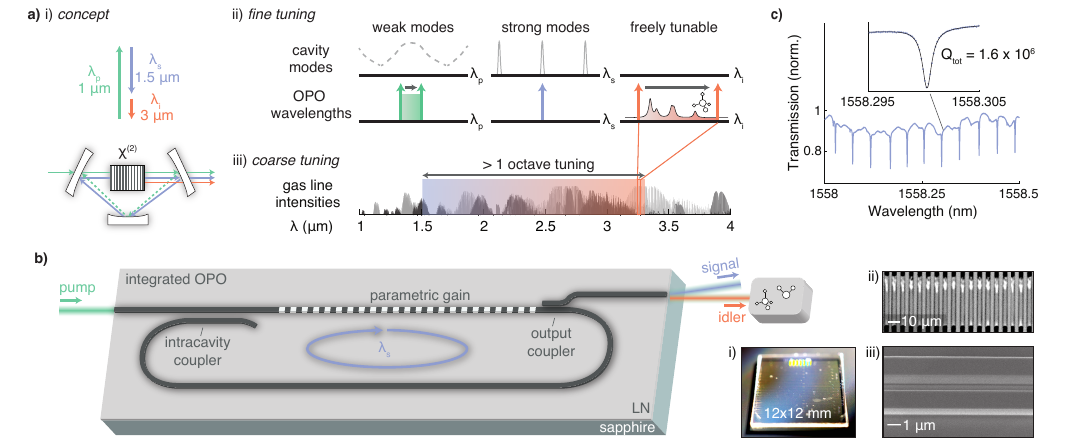}
  \end{center}
 \caption{\textbf{OPO concept and design.} \textbf{(a)} Inset (i): Diagram of a pump-enhanced SRO operating at $1$/$1.5$/$3$~µm. Inset (ii): Cavity designed to strongly resonate the signal, weakly resonate the pump, and leave the idler free to tune continuously for MIR spectroscopy. Inset (iii): Wide coarse tunability of single integrated OPO overlayed on spectral line intensities (log scale) for CH$_4$, H$_2$O, N$_2$O, CO$_2$, and NH$_3$ vs. wavelength. \textbf{(b)} Schematic of the on-chip integrated pump-enhanced SRO on LN-on-sapphire photonics. Inset (i): Image of fabricated $12 \times 12$~mm chip containing 15 devices. Inset (ii): second harmonic generation microscopy image of periodically poled domains. Inset (iii): scanning electron micrograph of waveguide coupler.  \textbf{(c)} High-quality-factor resonant modes for signal wavelengths near $1550$~nm. Inset: Lorentzian fit for single mode at $1558.3$~nm.}
 \label{fig:fig1}
\end{figure*}

Because of the limitations of bulk systems, OPO miniaturization has been actively pursued. Well-established systems include integrated weakly-confining waveguide cavities \cite{hofmann_MidInfrared_2000}, polished crystals \cite{phillips_Continuous_2011}, and whispering-gallery resonators \cite{amiune_Pump_2022}. Moreover, recent nanofabrication breakthroughs have led to on-chip planar nanophotonic circuits in strongly nonlinear materials such as lithium niobate (LN). Sub-wavelength transverse mode confinement in these architectures allows enhanced nonlinear efficiency \cite{luo_Highly_2018, wang_Ultrahighefficiency_2018}, dispersion engineering for ultrabroadband operation \cite{jankowski_Ultrabroadband_2020, jankowski_Dispersionengineered_2021, ledezma_Intense_2022}, and capability for complex nonlinear photonic circuits \cite{stokowski_Integrated_2022}. As a result, the first on-chip OPOs integrated with highly-scalable, small-footprint, nanophotonic circuits have recently been developed \cite{mckenna_Ultralowpower_2022, lu_Ultralowthreshold_2021, roy_VisibletomidIR_2022, ledezma_Octavespanning_2022, perez_Highperformance_2023, lu_Kerr_2022}.

Despite these rapid advances, recent nanophotonic integrated OPOs thus far have limited capability for MIR spectroscopy. One reason for this is that established nonlinear integrated photonic platforms utilize a silica undercladding that becomes strongly absorptive past $3$~µm \cite{mishra_Midinfrared_2021}, limiting MIR performance. Another crucial reason is that engineering nanophotonic OPOs with sufficiently stable and precise tuning over fine spectroscopic lines is challenging. Bulk OPO-based spectroscopy systems usually achieve ideal tuning behavior by engineering the cavity in a singly-resonant configuration with a resonant signal wave and non-resonant, freely-tunable MIR idler wave \cite{vainio_Midinfrared_2016, breunig_Continuouswave_2011, henderson_Low_2006, lindsay_Continuouswave_2003, vainio_Singly_2008, bisson_Broadly_2001}. Developing such wavelength-selective behavior within a high-quality-factor nanophotonic cavity is difficult. This has led previous integrated OPOs to simultaneously resonate signal and idler beams in either doubly- \cite{roy_VisibletomidIR_2022, ledezma_Octavespanning_2022} or triply-resonant \cite{mckenna_Ultralowpower_2022, lu_Ultralowthreshold_2021, perez_Highperformance_2023, lu_Kerr_2022} configurations, creating complex tuning dynamics undesirable for spectroscopy. 

Here we demonstrate an efficient, broadly-tunable, continuous-wave integrated MIR OPO and use it for gas spectroscopy. This single-wavelength MIR source complements broadband integrated MIR frequency comb sources \cite{bao_Architecture_2021, shams-ansari_Thinfilm_2022, yu_Siliconchipbased_2018, grassani_Mid_2019} that can exhibit more complex dynamics, difficult calibration/stabilization, low efficiency, and limited resolution. Pumped with continuous-wave light at $\lambda_p=1$~µm, a single dispersion-engineered device exhibits broad tuning over an octave from $1.5\text{--}3.3$~µm. By engineering a wavelength-selective, high-quality-factor cavity, we realize pump-enhanced singly-resonant MIR OPO operation. The OPO's reliable tuning behavior allows us to measure the spectra of methane and ammonia, demonstrating the spectrosopic potential of OPOs within a fully-chip-integrated platform. We discuss clear paths towards further enhancing the current OPO for widespread, practical use by improving overall system efficiency, near-degenerate performance, and gap-free tuning range.

\section{Results}
\subsection{Device concept and operation}
Fig.~\ref{fig:fig1}a illustrates our OPO design concept. An optical cavity incorporates a $\chi^{(2)}$ nonlinear crystal that provides parametric amplification between $\lambda_p = 1$~µm pump light and generated signal/idler light at $\lambda_s = 1.5$~µm and $\lambda_i = 3$~µm (Fig.~\ref{fig:fig1}a.i). We design the cavity to be strongly resonant for $\lambda_s$, weakly resonant for $\lambda_p$, and non-resonant for $\lambda_i$, classifying it as a pump-enhanced singly-resonant OPO (SRO) \cite{lindsay_Continuouswave_2003}. This design allows the MIR idler to freely tune for spectroscopy. An effective, simple SRO fine tuning method \cite{vainio_Midinfrared_2016, breunig_Continuouswave_2011, henderson_Low_2006, vanherpen_Tuning_2002} sweeps $\lambda_p$ while $\lambda_s$ clamps on a strong cavity resonance, so $\lambda_i$ tunes freely by energy conservation, \textit{e.g.}, over molecular absorption peaks (Fig.~\ref{fig:fig1}a.ii). Tuning the temperature and pump wavelength broadly adjusts the OPO output over $1.5\text{--}3.3$~µm (Fig.~\ref{fig:fig1}a.iii), which overlaps fundamental vibrational transitions of dozens of small molecules (\textit{e.g.} CO$_2$, CH$_4$, H$_2$O, and NH$_3$) important for spectroscopic monitoring.

We implement the integrated OPO device (Fig.~\ref{fig:fig1}b) in a photonic circuit composed of etched LN-on-sapphire ridge waveguides. Deeply-etched LN-on-sapphire photonics, with substrate transparency up to $4.5$~µm, have enabled dispersion-engineered broadband MIR generation up to $4$~µm \cite{mishra_Midinfrared_2021, mishra_Ultrabroadband_2022}. We fabricate $15$ OPOs with different design parameters on a $12\times12$~mm LN-on-sapphire chip (Fig.~\ref{fig:fig1}b.i), then focus on the optimal device for the experiment. Periodically poling one of the LN waveguides (Fig.~\ref{fig:fig1}b.ii) compensates for phase-velocity mismatch and allows broadband parametric gain. We choose the parametric gain waveguide geometry ($878$~nm LN film, $600$~nm etch, and $1.95$~µm top width) to enable strong fundamental transverse electric mode confinement at pump/signal/idler wavelengths (Ext Fig.~\ref{fig:DevDesign}a) and large parametric gain from modal overlap. Moreover, choosing this geometry produces ultrabroadband gain at degeneracy resulting from near-zero signal/idler group velocity dispersion (GVD) (Sec. \ref{sec:tunability} and Methods). 

\begin{figure*}[ht]
  \begin{center}
      \includegraphics[width=\textwidth]{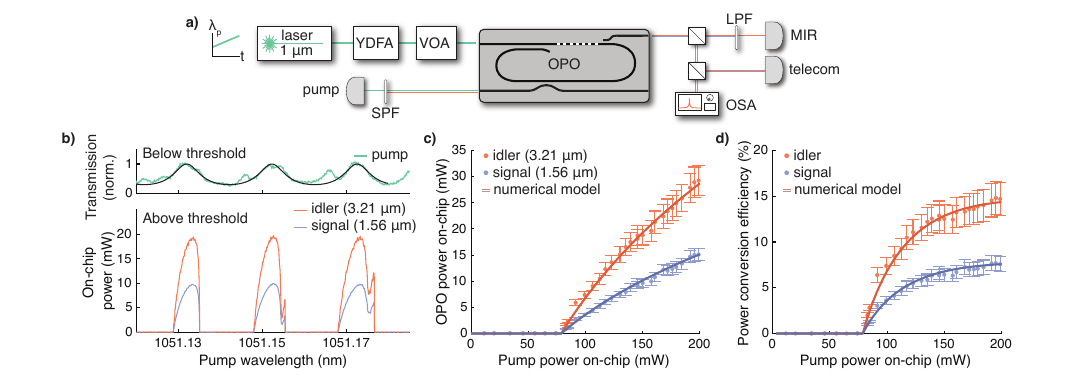}
  \end{center}
 \caption{\textbf{OPO power and efficiency.} \textbf{(a)} Simplified diagram of measurement setup for power and spectral characterization (full diagram in Ext. Fig.~\protect\ref{fig:MeasSetup}). YDFA: Ytterbium-doped fiber amplifier; VOA: variable optical attenuator; LPF: long-pass filter; SPF: short-pass filter; OSA: optical spectrum analyzer. \textbf{(b)} Top panel: Normalized pump transmission below threshold ($P_{pump}=45 \text{ mW}$) reveals low-finesse pump resonances with Airy function fit to a finesse of 2.4 (see Methods). Bottom panel: OPO output powers versus pump wavelength, at $P_{pump} = 157 \text{ mW}$ on-chip. \textbf{(c,d)} \textbf{(c)} Output on-chip powers and \textbf{(d)} power conversion efficiency vs. pump power. Error bars come from uncertainty in fiber-chip and chip-detector collection efficiency (see Methods). Solid lines from numerical modeling of the pump-enhanced SRO.}
 \label{fig:fig2}
\end{figure*}

The pump-enhanced SRO cavity combines waveguide bends with two crucial engineered elements: the output coupler and intracavity coupler. The output coupler (Fig.~\ref{fig:fig1}b.iii) is a directional coupler designed for ${\sim}100\%$ transfer of MIR light out of the cavity while only extracting ${\sim}1\%$ of telecom light. The intracavity coupler is an adiabatic coupler designed for broadband, ${\sim}100\%$ transfer of telecom-wavelength light to enable strong signal resonances. 
To verify the strong cavity modes at $\lambda_s$, we sweep resonances with a tunable telecom laser (Ext. Fig.~\ref{fig:LinNonlin}a), revealing sharp, low-loss  signal modes with total quality factor $Q_{tot} = 1.3-1.6 \times 10^6$ (Fig.~\ref{fig:fig1}c). This corresponds to ${\approx}12\%$ round-trip loss in the $22.3$~mm-length cavity. Extracted intrinsic/extrinsic $Q$-factors for the undercoupled cavity are $Q_i = 1.35-1.7 \times 10^6$ and $Q_{ex} \approx 20 \times 10^6$, respectively. High  $Q$-factors extend over our telecom laser's whole tuning range ($1500\text{--}1640$~nm, Ext. Fig.~\ref{fig:LinNonlin}b). Meanwhile, the $1$-µm pump only weakly resonates, with cavity finesse $F=2.5\text{--}3.5$, corresponding to ${\approx}11\%$ total power recirculation and $2\times$ intracavity power enhancement (Ext. Fig.~\ref{fig:PumpRes}).

To operate the device, we couple continuous-wave pump light onto the chip using a lensed fiber with ($33 \pm 2$)\% coupling efficiency (Methods). When parametric gain provided by the pump exceeds round-trip signal loss, the device oscillates, generating signal and idler photons. We collect output light with a multimode fiber (${\sim}3$\% MIR chip-to-fiber collection efficiency, Methods) and use the idler beam for MIR spectroscopy (Fig.~\ref{fig:fig1}b). We attribute the few-percent chip-to-fiber collection efficiency to the roughly-cleaved output fiber facet and mismatch between high-NA LN waveguide and $\text{NA} = 0.2$ fiber. Chip-fiber and fiber-chip coupling efficiencies could be improved dramatically to $80\text{--}90\%$ using cladding mode-matching waveguides \cite{hu_Highefficient_2021, li_Edge_2022, zhu_Ultrabroadband_2016} and/or placing a high-NA lens on the output (${>}70\text{--}80\%$ efficiency measured on a different chip/setup).

\subsection{Power characterization}
\label{sec:OpPowEff}

\begin{figure*}[t]
  \begin{center}
    \includegraphics[width=\textwidth]{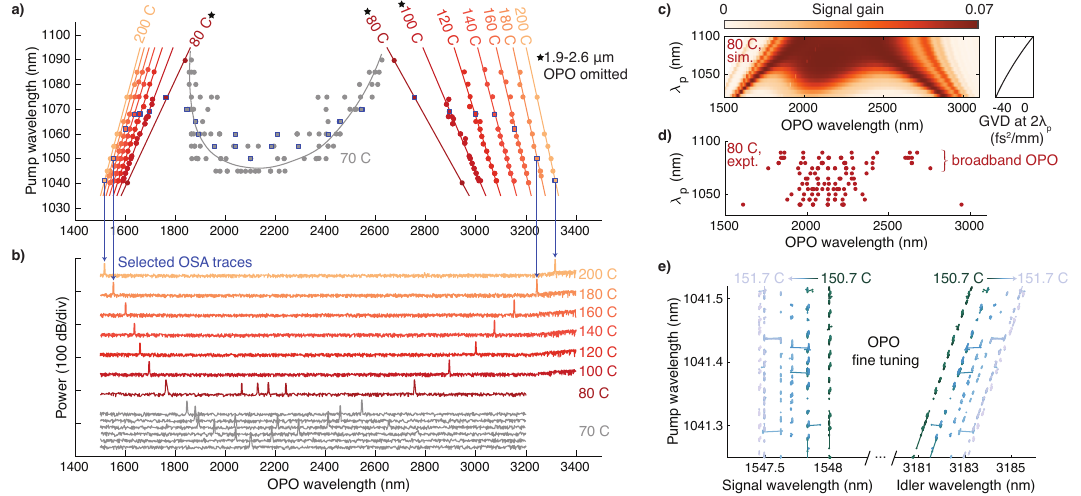}
    \end{center}
   \caption{\textbf{Coarse and fine tunability.} \textbf{(a)} OPO wavelength tuning curves vs. $\lambda_p$ for different temperatures. At $80$ and $100$~\textdegree C, the near-degenerate OPO from $1.8\text{--}2.6$~µm is omitted for clarity (complete data shown in Ext. Fig.~\protect\ref{fig:NearDegen}e). At $80$~\textdegree C and above, data are fit to lines, and at $70$~\textdegree C the U-shaped curve is a guide to the eye (simulated tuning in Ext. Fig.~\protect\ref{fig:NearDegen}a). Points marked with a blue outline have their output optical spectra plotted in \textbf{(b)}. \textbf{(c,d)} Broadband, $1.3$~µm-spanning OPO output observed at a single temperature ($80$~\textdegree C). \textbf{(c)} OPA gain spectrum simulation, showing degenerate OPO gain bandwidth increasing as $\lambda_p$ increases and group-velocity dispersion (GVD) at degeneracy ($2\lambda_p$) approaches zero. \textbf{(d)} OPO spectral peaks recorded in experiment agree with simulation and exhibit $1$~µm OPO span for $\lambda_p$ near $1075$~nm. \textbf{(e)} OPO fine tuning operation: for small changes in $\lambda_p$, $\lambda_s$ stays fixed without excessive mode hops while $\lambda_i$ increases monotonically, allowing for near-continuous spectral coverage. Gaps in the spectra are due to the weak pump wavelength resonance.}
  
  \label{fig:fig3}
\end{figure*}

Utilizing the characterization setup in Fig. \ref{fig:fig2}a, we tune the device to $170$~\textdegree C and pump near $\lambda_p=1.051$~µm to obtain clean non-degenerate parametric oscillation at $\lambda_s=1.56$~µm, $\lambda_i=3.21$~µm. Because of weak pump resonances (Fig.~\ref{fig:fig2}b, top), intracavity pump intensity and hence generated signal/idler output (Fig.~\ref{fig:fig2}b, bottom) varies periodically with $\lambda_p$. Because the pump resonance is weak, tuning to a specific $\lambda_p$ leads to stable continuous-wave oscillation for ${>}10\text{--}15$ minutes without any cavity or laser stabilization (Ext Fig.~\ref{fig:TimeStability}). 

We then scan $\lambda_p$ for different pump powers and record maximum generated signal/idler power. We observe clear pump depletion but do not precisely quantify it due to background pump light scattering into the multimode collection fiber. The device begins oscillating with $80 \pm 6$ mW on-chip threshold pump power (Fig.~\ref{fig:fig2}c). Above threshold, the generated signal/idler powers monotonically increase with pump power. With ${\sim}200$ mW on-chip pump power, the device produces a maximum of $29 \pm 3$ mW on-chip power at 3.2~µm. This power level has been used for portable sensor systems \cite{sgobba_Compact_2022} and exceeds the typical required power for shot-noise-limited MIR detection (${\sim}0.1$ mW) \cite{vainio_Midinfrared_2016}. The on-chip power conversion efficiency of signal/idler also increases monotonically within the range of pump power sweep (Fig.~\ref{fig:fig2}d). We measure a maximum of ($15 \pm 2$)\% on-chip power conversion efficiency ($45\%$ quantum efficiency) from pump to MIR idler. An ideal OPO produces nearly $100\%$ quantum efficient conversion ($\approx 33\%$ power conversion at these wavelengths) \cite{breunig_Continuouswave_2011}. Our device's deviation is likely caused by modal/radiative scattering of pump light in waveguide tapers (see Methods), MIR losses from \textit{e.g.} surface-adsorbed molecules \cite{mishra_Midinfrared_2021, mishra_Ultrabroadband_2022}, and inefficient MIR light transfer in the output coupler. 

The measured dependence of the emitted MIR light on input pump power aligns well with numerical modeling of a weakly pump-enhanced SRO (Fig.~\ref{fig:fig2}c-d, solid lines), verifying that the device behaves as designed. In our modeling (Methods) we assume the measured values of total $Q$-factor ($1.6$ million), pump recirculation ($11\%$, Methods), and normalized efficiency ($41 \%/(\text{W}\cdot\text{cm}^2)$). The numerically-modeled on-chip idler output powers are scaled by $0.46$ to account for the effective MIR extraction efficiency, and intracavity signal powers are scaled by $0.013$ to account for the intended small (${\sim}1\%$) signal extraction from the cavity. 

\subsection{Tunability}
\label{sec:tunability}

\subsubsection{Coarse tunability}

We tune our OPO's output wavelength over an octave of bandwidth using a combination of temperature and pump wavelength (Fig.~\ref{fig:fig3}a,b). At the higher temperatures of $100\text{--}200$~\textdegree C we access the ``far-from-degenerate'' regime with widely-separated signal and idler  ($\lambda_s=1.5\text{--}1.7$~µm, $\lambda_i=3\text{--}3.3$~µm) (Fig.~\ref{fig:fig3}a). We observe sufficiently reliable tuning for spectroscopy at these operating temperatures and clean output spectra (Fig.~\ref{fig:fig3}b). In this regime, we measure MIR output wavelengths up to $3.315$~µm at $200$~\textdegree C, limited by the temperature control range and pump amplifier bandwidth. The high operation temperature is only due to phase matching in this device; future devices can extend deeper into the MIR at lower temperatures by lithographically defining a different poling period. In our device, lower temperatures from $70\text{--}90$~\textdegree C access the ``near-degenerate'' regime ($1.7 \text{ µm} < \lambda_s, \lambda_i < 2.7 \text{ µm}$), exhibiting broad bandwidths and tunability but also some complex multimoded behavior. From $80\text{--}100$~\textdegree C, the OPO sometimes oscillates simultaneously in the near-degenerate and far-from-degenerate regimes.

Pump wavelength tuning at a fixed temperature tunes the device reliably and rapidly over a large range (Fig.~\ref{fig:fig3}a). From $80\text{--}200$~\textdegree C, the ${>}2.8$~µm idler tunes roughly linearly with pump wavelength. The fitted tuning slope $d\lambda_i/d\lambda_p\approx -2$ at higher temperatures and increases to $-4.2$ at $100$~\textdegree C. This equates to $100\text{--}200$~nm MIR wavelength tuning at a given temperature with $50$~nm of pump tuning.

As we further decrease the temperature, wavelength tunability rapidly increases as the device begins oscillating at near-degenerate signal/idler waveguide modes with near-zero GVD. At $70$~\textdegree C, we operate the device in the anomalous dispersion regime, resulting in a U-shaped tuning curve that spans over $800$ nm (Fig.~\ref{fig:fig3}a) and agrees with simulations (Ext. Fig.~\ref{fig:NearDegen}a). Near degeneracy, the accessible gain bandwidth broadens from cancellation of odd-order dispersion, allowing oscillation at multiple different signal/idler pairs (Fig.~\ref{fig:fig3}b). At $80$~\textdegree C, the device operates near the signal/idler zero-GVD point, resulting in broadband OPO output spanning $1.3$~µm, a bandwidth approaching a full octave, at a single temperature (Fig.~\ref{fig:fig3}c,d). The increasingly broadband near-degenerate OPO as we increase $\lambda_p$ and approach zero-GVD at 2$\lambda_p$ agrees well with simulation (Fig.~\ref{fig:fig3}c) \cite{lin_Optical_2020, muraviev_Massively_2018}. From $\lambda_p\approx1075\text{--}1090$~nm, the single-device, single-temperature, OPO output spans $1.7\text{--}2.7$~µm. This 65~THz-spanning ultrabroadband gain bandwidth matches that of state-of-the-art pulsed-pump dispersion-engineered thin-film-LN parametric amplifiers \cite{ledezma_Intense_2022, jankowski_Quasistatic_2022}. To fully harness the broadband OPO operation, future devices could employ an on-chip wavelength control element (e.g. \cite{wang_Photoniccircuitintegrated_2023, stern_Batteryoperated_2018}) rapidly tunable using LN's electro-optic effect and selective of a particular oscillating mode. Full near-degenerate dispersion-engineering details are described in Methods.

\subsubsection{Fine tunability}

We finely tune the SRO's MIR emission wavelength with sufficient control for use in spectroscopy. At a fixed temperature, we tune the pump laser wavelength. For small changes of $\lambda_p$, $\lambda_s$ stays approximately constant without excessive mode hops while the MIR $\lambda_i$ tunes by energy conservation (Fig.~\ref{fig:fig3}e). The vertical gaps visible in these fine tuning curves are caused by weak pump resonance enhancement, not signal mode hops. Typical gap-free tuning range is $60\text{--}80$~pm at $3184$~nm ($1.8\text{--}2.4$~GHz), reflecting that the OPO is activated for around one-third of the pump cavity FSR ($5.5$~GHz, Fig.~\ref{fig:fig2}a). Eliminating the weak pump resonance in an optimized fully-singly-resonant cavity design will allow broader gap-free tuning range. Despite this discontinuous tuning at a fixed temperature, adjusting the chip temperature by only $1$~\textdegree C results in nearly uniform MIR wavelength coverage as different signal modes are selected to oscillate. The signal mode hops when $\lambda_p$ is detuned sufficiently large amounts (Methods, Ext. Fig.~\ref{fig:FineTun}a).

\subsection{Proof-of-concept spectroscopy}
\begin{figure*}[t]
  \begin{center}
      \includegraphics[width=\textwidth]{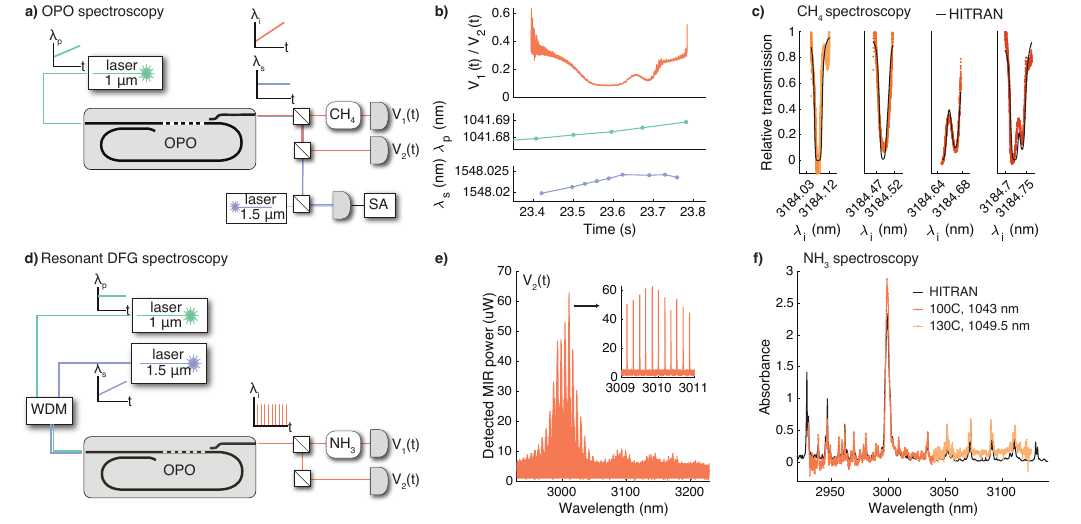}
  \end{center}
 \caption{\textbf{MIR spectroscopic operation.}  \textbf{(a-c)} OPO spectroscopy of narrow methane lineshapes. \textbf{(a)} Experimental setup showing scanning pump laser input. SA: RF spectrum analyzer. \textbf{(b)} Example trace of measured absorption signal, pump wavelength, and signal wavelength versus time. \textbf{(c)} Methane Doppler-broadened transmission lineshapes for four different scans, compared with data from HITRAN database \protect\cite{gordon_HITRAN2020_2022}. \textbf{(d-f)} Subthreshold resonant DFG spectroscopy for rapid spectroscopy of broad lineshapes over ${>}100$~nm MIR bandwidth. \textbf{(d)} Experimental diagram depicting fixed pump laser input and scanning telecom signal laser. WDM: wavelength division multiplexer. \textbf{(e)} Example trace of output generated MIR signal at detector $2$ vs. MIR wavelength. \textbf{(f)} Detected MIR absorbance spectrum of ammonia, compared with data from HITRAN database \protect\cite{gordon_HITRAN2020_2022}.}
 \label{fig:fig4}
\end{figure*}
\subsubsection{OPO spectroscopy of methane}
 
We direct part of the output idler light to a low-pressure ($20$~Torr) methane gas cell (Fig.~\ref{fig:fig4}a) to measure its absorption spectrum. Tuning the device to $151$~\textdegree C with $\lambda_p$ $\approx 1041 \text{~nm}$ shifts the OPO idler output to a cluster of methane absorption lines at $3184$~nm. To sweep the generated MIR output over the methane lines, we sweep $\lambda_p$ in a narrow range (Fig.~\ref{fig:fig4}a). During this measurement, the OPO output signal wavelength $\lambda_s$ stays nearly constant, while the idler wavelength $\lambda_i$ increases with time (Fig.~\ref{fig:fig4}a). The portion of the MIR light passing through the methane cell couples into a photodiode, generating the voltage signal $V_1(t)$. The reference beam of MIR light couples into a second photodiode, generating a reference signal $V_2(t)$. We plot an example trace of the unprocessed relative gas cell transmission scan $V_1(t)/V_2(t)$ in Fig.~\ref{fig:fig4}b, which scans over two absorption peaks.

To calibrate the wavelength axis $\lambda_i(t)$ of the swept MIR beam, we measure $\lambda_p(t)$ and $\lambda_s(t)$ and infer $\lambda_i(t)$ by energy conservation. This method allows us to use precise and more readily available near-IR wavelength measurement tools to infer MIR emission properties. We measure $\lambda_p(t)$ with a wavemeter in the input path (Fig.~\ref{fig:fig4}b). Meanwhile, $\lambda_s(t)$ is measured by beating a portion of the generated signal beam against a reference laser on a fast photodiode. The beatnote is read in an RF spectrum analyzer, from which we extract $\lambda_s(t)$ (Fig.~\ref{fig:fig4}b). The observed small ($5$~pm) redshift of $\lambda_s(t)$ is much smaller than the cavity free spectral range (${\sim}45$~pm), indicating that the signal mode does not mode hop during this scan, but only shifts slightly, likely due to heating at higher OPO power.

We tune the device to four different absorption transitions of methane near $3184$~nm and collect spectra (Fig.~\ref{fig:fig4}c). After background subtraction (see Methods), collected experimental spectra agree well with HITRAN reference curves \cite{gordon_HITRAN2020_2022}. The clean, stable MIR OPO output easily resolves the low-pressure, Doppler-broadened methane peaks with linewidths down to $10$~pm/$300$~MHz/$.01$~cm$^{-1}$. This spectral resolution highlights an advantageous aspect of the widely-tunable single-wavelength integrated source compared to an integrated frequency comb, which in integrated incarnations have few~GHz$\text{--}100$~GHz resolution limited by the cavity free spectral range \cite{bao_Architecture_2021, hu_Highefficiency_2022, shams-ansari_Thinfilm_2022, yu_Siliconchipbased_2018, grassani_Mid_2019, wang_Midinfrared_2013}.

\subsubsection{Resonant DFG spectroscopy of ammonia}

In addition to operating as an OPO, the broadband operation and singly-resonant nature of our device makes it attractive as a source of MIR light generated by difference frequency generation (DFG). Here, we pump the OPO cavity below threshold, now leaving $\lambda_p(t)$ constant over time (Fig.~\ref{fig:fig4}d). We instead seed the device with a scanning telecom-band laser $\lambda_s(t)$. Injected seed builds up strongly when $\lambda_s(t)$ matches a signal cavity resonance and generates a bright MIR idler beam by DFG. Hence, $\lambda_i(t)$ consists of discretely-spaced MIR peaks (Fig.~\ref{fig:fig4}d). In our device's SRO cavity, peaks at $\lambda_i$ will be equally spaced in frequency at the signal cavity FSR ($\approx$5.6~GHz) over the entire gain bandwidth. By contrast, in a doubly- or triply-resonant device, generated MIR peaks would be much more sparse because of the requirement of simultaneous signal and idler resonance. The wide availability of rapidly tunable telecom-band lasers, including on-chip and LN-integrated devices \cite{tran_Tutorial_2019, li_Integrated_2022, opdebeeck_III_2021}, makes this resonant DFG technique highly accessible.

We demonstrate the broadband resonant DFG spectroscopy by detecting atmospheric pressure ammonia, which exhibits broad lineshapes with $0.5\text{--}10$~nm peak widths. As in the methane experiment, the MIR output splits into a gas cell and reference path.  We measure $\lambda_p$ with a wavemeter, assume $\lambda_s(t)$ sweeps linearly with time, then infer $\lambda_i(t)$ by energy conservation. Fig.~\ref{fig:fig4}e shows a typical trace of discrete MIR peaks at detector $2$ vs. $\lambda_i(t)$, where $\lambda_p = 1043$~nm and $\lambda_s(t)$ sweeps from $1535$ to $1620$~nm. Generated equally-spaced MIR lines (Fig.~\ref{fig:fig4}e, inset) are strong over an ${\sim}100$~nm bandwidth (equates to sweeping $\lambda_s$ only $30$~nm), and the ${>}10$~µW MIR output can be detected directly by a DC-coupled photodiode. 

By dividing the signal path's discrete peak heights by those from the reference path, we obtain broadband spectra of ammonia with $5$~GHz resolution (Fig.~\ref{fig:fig4}f). The presented data consists of two scans, each with high signal-to-noise ratio over $100$~nm MIR bandwidth. Adjusting temperature and $\lambda_p$ tunes the center wavelength of the two scans exactly as the OPO is coarsely tuned (Fig.~\ref{fig:fig3}b). We resolve ammonia's narrower features with ${\sim}0.5$~nm peak width alongside broader 10~nm peaks in agreement with the HITRAN database.

\section{Discussion}
In summary, we have designed and implemented an integrated nanophotonic OPO and demonstrated operation for MIR spectroscopy. Such a device inherits the useful advantages of bulk OPOs as MIR spectroscopic light sources (widely-available near-IR laser pumps, high efficiency, broad tunability, and high resolution) while adding the benefits of nanophotonic integration (reduced footprint, better stability, lower threshold powers, broadband operation via dispersion engineering, and integration capability). The key enabling advance here is the fabrication of high-quality factor, wavelength-selective cavities built from the MIR-compatible LN-on-sapphire platform. With the miniaturization of such a useful MIR spectroscopic technology onto a fully-chip-integrated platform, a plethora of applications can be envisaged, from deployable gas monitoring systems to portable, real-time MIR biosensors. 

Our work outlines a clear path for improving the device sufficiently to realize powerful and deployable sensors. As highlighted in the text, including an electro-optically-tunable wavelength-selective intracavity etalon would allow precise, rapid, and low-power control over the broad demonstrated gain bandwidths. In addition, further gains in efficiency are important and within reach. These will come from improvements in input fiber-to-chip and output chip-to-detector coupling efficiencies. Simulations show that utilizing cladding mode matching waveguides and/or free space optics would raise edge coupling efficiencies to ${>}70\%$. Moreover, the simulated normalized efficiency is ${\sim}7\times$ larger than the experimentally-obtained value ($43 \text{ } \%/(\text{W}\cdot\text{cm}^2)$), likely due to fabrication imperfections preventing coherent nonlinear enhancement over the full waveguide length. By improving waveguide fabrication we expect threshold powers as low as ${\sim}10$ mW, within the output range of heterogeneously integrated lasers near $1$ \textmu m \cite{tran_Extending_2022} and thus potentially enabling full pump-OPO on-chip integration. 

\bibliography{apssamp}

\section{Methods}
\renewcommand{\figurename}{Extended Data Fig.}
\setcounter{figure}{0}
\subsection{Coupler, bend, and taper design details}
\label{sec:CoupBendTaper}

\subsubsection{Intracavity coupler}
The intracavity coupler (Ext. Fig.~\ref{fig:DevDesign}b) is an adiabatic coupler designed to weakly couple pump light at $1$~µm and strongly couple signal light at ${>}1.5$~µm. The coupler was designed using local coupled mode theory simulations of the slow transfer of light from the waveguide emerging from the resonator bend to the poled section waveguide. We utilize a symmetric adiabatic waveguide coupler where two neighboring waveguides of width $0.7$/$1.0$~µm are tapered to widths $1.0$/$0.7$~µm width, respectively, over $1$~mm length. In order to maximize the adiabatic transition near the degenerate point where both waveguides have equal width ($0.85$~µm), the coupler is divided into three sections: two $150$~µm-length fast-tapered couplers at the beginning and end of the coupler and a slowly-varying $700$~µm section in the middle of the coupler. The slowly-varying middle section accounts for $20\%$ of the total waveguide width change, while the fast-varying sections account for the remaining $80\%$.

\subsubsection{Output and diagnostic couplers}
\begin{figure*}[t]
  \begin{center}
      \includegraphics[width=\textwidth]{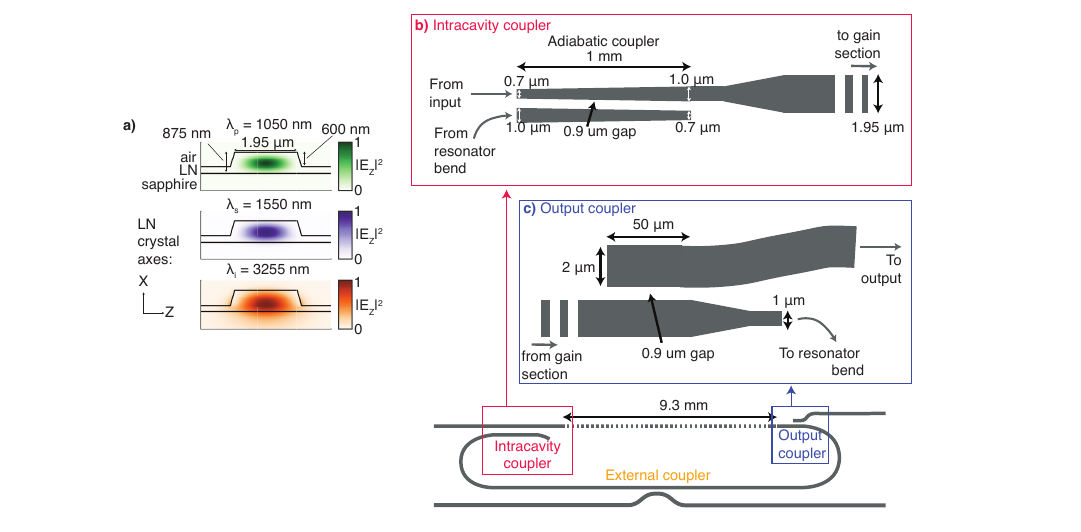}
  \end{center}
 \caption{\textbf{Device basics in more detail.} \textbf{(a)} Simulated normalized fundamental transverse electric (TE0) modal profiles $|E_Z|^2$ for $\lambda_p$~=~1050~nm, $\lambda_s$~=~1550~nm, and $\lambda_i$~=~3255~nm, along with annotations of waveguide top width, film thickness, and etch depth. \textbf{(b,c)} Detailed diagram (not to scale) of resonator design, focusing on the region around the \textbf{(a)} intracavity adiabatic coupler and \textbf{(b)} output coupler. }
 \label{fig:DevDesign}
\end{figure*}

The output coupler (Ext. Fig.~\ref{fig:DevDesign}c) is a simple directional coupler consisting of two identical $2$~µm-width waveguides separated by a $0.9$~µm coupling gap over a length of $50$~µm. The diagnostic coupler (Ext. Fig.~\ref{fig:DevDesign}) uses the same coupling geometry as the output coupler in order to obtain $\sim$1\% coupling of telecom light in/out of the cavity for measuring the cavity resonances (Fig.~\ref{fig:fig1}c).

\subsubsection{Resonator bends}
Following the poled section and the output coupler, the waveguides are tapered to $1$~µm width to ensure that the resonator is single-moded at telecom wavelengths for cleaner mode structure. This also effectively filters out MIR light ${>}3$~µm that cannot be well-confined in the smaller waveguide. The waveguide bends are Euler bends \cite{vogelbacher_Analysis_2019}.

\subsubsection{Waveguide tapers}
Pump light incident onto the chip edge couples into a $1.7$~µm-width waveguide, then tapers down to $0.7$~µm as it reaches the adiabatic coupler. After the adiabatic coupler, the light is confined in a $1$~µm-width waveguide before it tapers up to the $1.95$~µm-width periodically-poled 9.3 mm gain section. At each of these tapers, fundamental TE mode pump light can be scattered into other modes or free space, but the exact loss rate cannot be extracted from the current chip.

\subsection{Device fabrication}
Device fabrication starts with a commercial MgO-doped, x-cut LN film on a c-cut sapphire substrate (NGK Inc.). The LN film is thinned using an ion mill, then poling electrodes (poling period $\Lambda=6.72$~µm) are patterned using electron-beam lithography and Cr metal liftoff. The LN is poled using high-voltage pulses (${\sim}900$~V), then poled domains are monitored using second-harmonic-generation microscopy. Electrodes are stripped using Cr etchant. Waveguides are patterned with electron-beam lithography (JEOL 6300FS 100 kV) and HSQ FOX-16 resist followed by argon ion mill etching (Intlvac). Finally, the chip is laser stealth diced to create clean edge facets for light in/out coupling.

\subsection{Measurement setup and calibrations}
\subsubsection{General setup}
\label{sec:MeasSetupGeneral}
\begin{figure*}[t]
  \begin{center}
      \includegraphics[width=\textwidth]{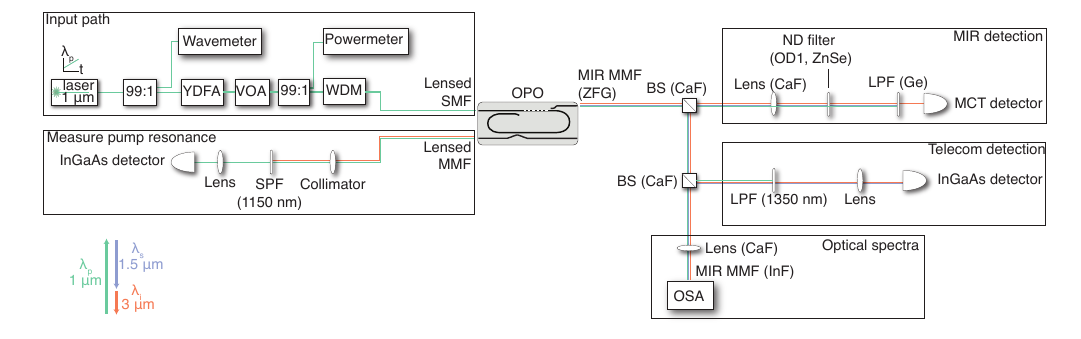}
  \end{center}
 \caption{\textbf{General measurement setup.} Abbreviations: YDFA - Ytterbium doped fiber amplifier; VOA - variable optical attenuator; WDM - wavelength division multiplexer; SMF - single-mode fiber; MMF - multimode fiber; SPF - short pass filter; LPF - long pass filter; ZFG - Zinc fluoride glass; BS - beamsplitter; OSA - optical spectrum analyzer.}
 \label{fig:MeasSetup}
\end{figure*}

A block diagram of the general setup used for measurements is shown in Ext. Fig.~\ref{fig:MeasSetup}. Our pump light source is a tunable external cavity diode laser (Toptica DL Pro). The laser can tune coarsely from $1010\text{--}1100$~nm with $0.1$~nm resolution, and finely using a voltage-controlled piezo within $40$~GHz. The laser output is fiber coupled, then routed through a $99$:$1$ splitter where the $1\%$ tap is sent to a near-IR wavemeter (Bristol Instruments Model $621$). Light from the $99\%$ port is amplilfied in a ytterbium-doped fiber amplifier with $1040\text{--}1090$~nm operating bandwidth (Civil Laser), then sent to a variable optical attenuator (OZ Optics). To calibrate power sent to the chip, $1\%$ of the light is tapped off to a powermeter (Newport), then the rest is sent to a $1$/$1.5$~µm wavelength division multiplexer, then into a lensed Hi1060 single mode fiber that couples light to the OPO device. 

The amount of intracavity pump light is measured by collecting the light exiting the bottom bus waveguide on the left of Ext. Fig.~\ref{fig:MeasSetup} with a lensed multimode silica fiber. The light is then sent through a fiber collimator, and a short-pass filter to remove $1.5$-µm output generated by the OPO, and focused onto an InGaAs detector. 

The output at telecom and MIR wavelengths is collected using a flat-cleaved MIR-compatible multimode fiber (Zinc fluoride glass, La Verre Fluoré). The output light is split into several paths. To detect the MIR light, we focus with a CaF lens (Thorlabs) then through a ZnSe OD1 ND filter (Thorlabs) to avoid saturating the detector. Finally, the light passes through a longpass filter (Ge) so that only MIR light reaches the MCT detector (Thorlabs PDAVJ5). To detect telecom, we use a $1350$~nm longpass filter (Thorlabs) before focusing light onto an InGaAs detector (which does not detect MIR photons). Finally, a portion of the output light is coupled into an InF MIR-compatible multimode fiber (Thorlabs) sent into a Yokogawa Optical Spectrum Analyzer (AQ6376).

\subsubsection{Pump fiber-to-chip coupling efficiency}
 We couple pump light with wavelength $1046\text{--}1056$~nm  in/out of a straight waveguide using two lensed SMF fibers. By dividing the power collected from the output lensed fiber by the power sent to the input lensed fiber, we infer the pump power coupling efficiency per edge of $\tilde{\eta}_{p, fiber-to-chip} = (33 \pm 2) \%$. We assume here that pump propagation loss is negligible. The uncertainty in the pump fiber-chip coupling comes from ripples in the waveguide throughput observed as $\lambda_p$ is tuned from $1046\text{--}1056$~nm. We attribute the ripples to excitation of higher-order pump modes in the multimoded LN waveguide, because a simultaneous measurement of parametric gain (which depends only on fundamental TE0 mode power) during the same wavelength scan does not follow the same ripples. 

\subsubsection{Nonlinear efficiency}
\label{sec:DetermineEta}
\begin{figure*}[t]
  \begin{center}
      \includegraphics[width=\textwidth]{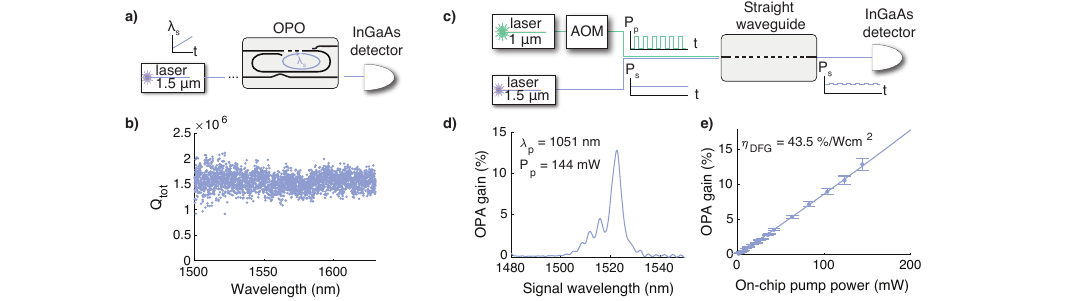}
  \end{center}
 \caption{\textbf{Basic linear and nonlinear device characterization.} \textbf{(a-b)} Resonator quality factor. \textbf{(a)} Diagram of apparatus used to determine resonator quality factor from linewidths of signal resonances, and \textbf{(b)} extracted total quality factor for $2621$ fitted modes ($174$ poor fits excluded from plot). \textbf{(c-e)} Nonlinear efficiency. \textbf{(c)} Diagram of measurement setup to extract normalized efficiency by optical parametric amplification (OPA) of the signal wave. \textbf{(d)} OPA gain versus signal wavelength for $144$ mW on-chip pump power at $1051$~nm. \textbf{(e)} Measured and fitted (Eq. \protect\ref{eq:siggain_cosh}) OPA gain vs. on-chip pump power.}
 \label{fig:LinNonlin}
\end{figure*}

The normalized efficiency of the periodically-poled gain section is calculated by measuring optical parametric amplification (OPA) on a straight waveguide adjacent to the OPO that was poled using the same electrodes (Ext. Fig.~\ref{fig:LinNonlin}c). In this experiment we couple both $1$~µm pump and $1.5$~µm signal onto the straight waveguide. We modulate the pump with a $10$~kHz square-wave using an acousto-optic modulator (Aerodiode). The periodic pump modulation periodically provides gain to the signal wave, which we measure with a lock-in amplifier. The relationship between measured signal gain and nonlinear efficiency can be derived starting from the coupled wave equations:
\begin{equation}
\begin{cases}
      \partial_z A_p = -\frac{\omega_p}{\omega_i} \sqrt{\eta_{DFG}} A_{s} A_{i} \\
      \partial_z A_{s} =  \frac{\omega_s}{\omega_i} \sqrt{\eta_{DFG}} A_{p} A_{i}^* \\
      \partial_z A_{i} =  \sqrt{\eta_{DFG}} A_p A_{s}^*,
\end{cases} 
\label{eq:CWE}
\end{equation}
where $A_{p,s,i}(z)$ are the power-normalized pump ($\omega_p$), signal ($\omega_s$), and idler ($\omega_i$) amplitudes with units of $\sqrt{\textrm{W}}$ and $\eta_{DFG}$ is defined as the normalized efficiency and with units of $\%/(\textrm{W}\cdot \textrm{cm}^2)$. For optical parametric amplification with an undepleted pump, these equations can be reduced to:
\begin{equation}
\begin{cases}
      \partial_z a_s = \gamma a_i^* \\
      \partial_z a_i = \gamma a_s^*, \\
\end{cases} 
\end{equation}
where $a_{s,i}$ are photon flux-normalized signal and idler amplitudes and $\gamma = -i\sqrt{\frac{\omega_s}{\omega_i}}\sqrt{\eta_{DFG}}A_p(0)$. The solution to this system is well-known \cite{jankowski_Dispersionengineered_2021}:
\begin{equation}
\label{eq:OPA_solution}
    \begin{cases}
    a_s(z) = \cosh(|\gamma|z) a_s(0) -i \sinh(|\gamma|z) a_i^*(0)\\
    a_i^*(z) = i\sinh(|\gamma|z) a_s(0) + \cosh(|\gamma|z) a_i^*(0),
    \end{cases}
\end{equation}
so with zero intial idler input ($a_i=0$) and fixed poling length $L_{pol}$, the telecom amplitude experiences the power gain:
\begin{align}
    \textrm{signal power gain} &\equiv  \frac{\left|a_s(L_{pol})\right|^2-\left|a_s(0)\right|^2}{\left|a_s(0)\right|^2}  \\
    \label{eq:siggain_cosh}
    &= \cosh^2(|\gamma|L_{pol}) - 1 \\
    & \approx \eta_{DFG}\left(\frac{\omega_s}{\omega_i}\right)P_p(0)L_{pol}^2
\label{eq:siggain_lowconv}
\end{align}
in the low gain limit. At a fixed $\lambda_p$ and pump power, we can sweep $\lambda_s$ and monitor the OPA gain (Ext. Fig.~\ref{fig:LinNonlin}d). The OPA gain is maximized when phase-matching is optimized. We then track the maximal phase-matched OPA gain as a function of $P_p(0)$ (Ext. Fig.~\ref{fig:LinNonlin}e). Fitting signal power gain as a function of pump power (using Eq. \ref{eq:siggain_cosh}) for known $L_{pol} = 0.93 \text{ cm}$ allows extraction of the nonlinear efficiency $\eta_{DFG} = 43.5 \text{ } \%/(\text{W}\cdot \text{cm}^2)$. The simulated normalized efficiency is around $300 \text{ } \%/(\text{W}\cdot \text{cm}^2)$.

\subsubsection{MIR collection efficiency for power sweep}

We calibrate the relationship between MIR on-chip power and detected voltage in the MIR MCT detector by simultaneously comparing OPA and difference-frequency-generation (DFG) processes in a straight nonlinear waveguide. For this calibration, we send a modulated pump ($10$~kHz square wave) along with a CW telecom signal wave onto a straight periodically-poled waveguide. The modulated pump produces parametric gain modulation in the telecom signal according to Eq. \ref{eq:siggain_lowconv}. Because of photon number conservation, the amplification of telecom photons is also accompanied by generation of the same number of MIR photons by DFG. The expected amount of detected MIR idler power is then:
\begin{align}
    P_{i,det} &= \tilde{\eta}_{i,chip-to-det} P_i(L_{pol}) \\
    &= \tilde{\eta}_{i,chip-to-det} \eta_{DFG}P_p(0)P_s(0)L_{pol}^2.
\label{eq:DFG_detected}
\end{align}
Dividing Eqs. \ref{eq:siggain_lowconv} and \ref{eq:DFG_detected} we can solve for the MIR chip-to-detector collection efficiency:
\begin{equation}
\tilde{\eta}_{i,chip-to-det} = \frac{P_{i,det}}{\text{signal power gain}\cdot (\omega_i/\omega_s)\cdot P_s(0)}.
\end{equation}
Using this method eliminates the contribution of any uncertainties in nonlinear efficiency $\eta_{DFG}$ and on-chip pump power. The uncertainties are dominated instead by $P_{i,det}$ and $P_{s}(0)$. 

With an on-chip pump power of $47.9$ mW and on-chip signal power of $P_s(0) = 0.8 \pm 0.06 \text{ mW}$, we measure a telecom $\text{signal gain} = 3.28 \%$. We also measure DFG idler of $P_{i,det}=17.8 \pm 1.1 \text{ nW}$. Combining these values leads to a MIR chip-to-detector collection efficiency of $\tilde{\eta}_{i,chip-to-det} = (0.16 \pm 0.016)$\%. Dividing out the attenuation from the OD1 filter and $50$:$50$ beamsplitter, this means that the MIR chip-to-fiber collection efficiency is around $3\%$.

\subsubsection{Telecom collection efficiency for power sweep}
For the telecom calibration, we first in-couple and out-couple $1550$~nm telecom light onto a straight waveguide using Hi1060 lensed fibers. By measuring the power sent into the input fiber and collected from the output fiber, we extract a telecom fiber-to-chip power coupling efficiency of $(30\% \pm 2) \%$. Then we switch the output telecom detection chain to that shown in Ext. Fig.~\ref{fig:MeasSetup}. By comparing the collected telecom power measured directly before the InGaAs telecom detector to the known on-chip power, we extract the telecom chip-to-detector collection efficiency of $(2 \pm 0.2)\%$, which is on the same order as that for MIR (previous section). We also calibrate the detector conversion efficiency to be $6.2$ V/mW, allowing for measured voltage to be converted to on-chip power.

\subsection{Pump resonance characterization}
\label{sec:PumpResChar}

\begin{figure*}[t]
  \begin{center}
      \includegraphics[width=\textwidth]{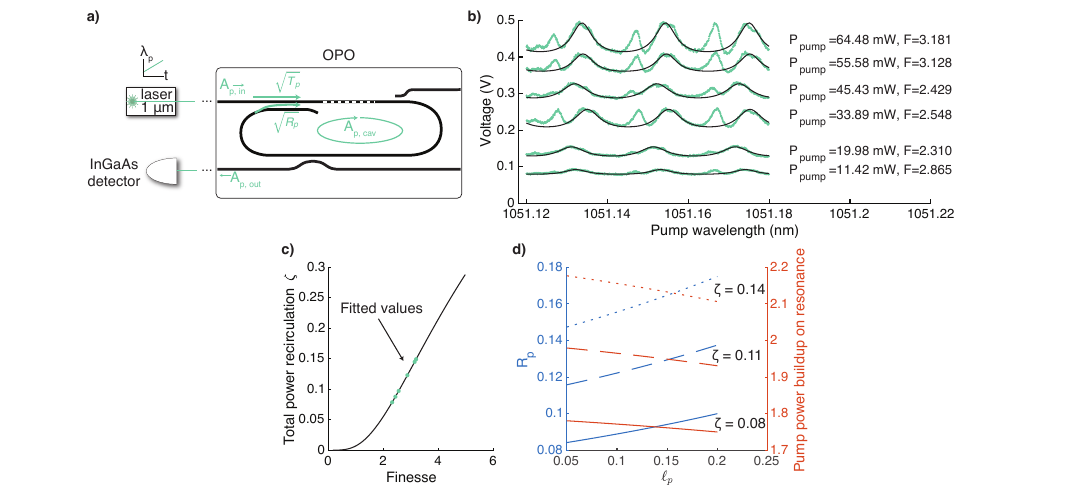}
  \end{center}
 \caption{\textbf{Weak pump resonances.} \textbf{(a)} Diagram of pump resonance measurement concept. $A_p$ denotes the power-normalized pump amplitude, and $R_p, T_p$ represent the cavity's reflection and transmission coefficients for $\lambda_p$ at the intracavity coupler. \textbf{(b)} Below-threshold pump transmission from the scans presented in Fig.~\ref{fig:fig2}, fitted to Airy functions and annotated with on-chip pump power and fitted cavity finesse. \textbf{(c)} Total power recirculation $\zeta$ of the pump as a function of fitted values of cavity finesse. \textbf{(d)} Extracted value of $R_p$ and on-resonant pump power buildup versus the assumed value of pump loss outside the intracavity coupler $\ell_p$. The solid, dashed, and dotted curves represent assumed values of $8\%$, $11\%$, and $14\%$ total pump recirculation respectively, corresponding to the range of values extracted in \textbf{(c)}.}
 \label{fig:PumpRes}
\end{figure*}

\begin{figure}[t]
  \begin{center}
      \includegraphics {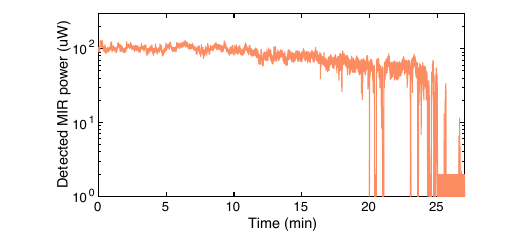}
  \end{center}
 \caption{\textbf{OPO time stability.} Typical OPO output MIR power vs. time, for pump wavelength of $1041.5$~nm and temperature around $150$~\textdegree C, corresponding to MIR wavelength output at $3.18$~µm. The OPO power output remains stable for around $10\text{--}15$ minutes without any active feedback before the low-finesse cavity resonance drifts too far from the pump wavelength.}
 \label{fig:TimeStability}
\end{figure}

Using the detection setup shown in Ext. Fig.~\ref{fig:PumpRes}a (more detail described in Ext. Fig.~\ref{fig:MeasSetup}), pump resonances are monitored for pump powers below the OPO threshold and are plotted in Ext. Fig.~\ref{fig:PumpRes}b. To fit these curves, we develop a simple model for the pump cavity.

In steady-state, the intracavity pump field $A_{p,cav}$ obeys the equation:
\begin{equation}
    A_{p,cav} = \sqrt{T_p}A_{p,in} + \sqrt{R_p}\sqrt{1-\ell_p}A_{p,cav}e^{-ikL}
\end{equation}
where $R_p$ is the pump power coupling ratio across the waveguide gap inside the intracavity coupler, $T_p = 1-R_p$ is the pump power transmission ratio of light that stays on the same waveguide through the coupler, $A_{p,in}$ is the pump amplitude on the input waveguide, $\ell_p$ is the round-trip power loss of the pump within the cavity excluding the intracavity coupler region, $k$ is the propagation constant of pump, and $L$ is the round-trip cavity length. Hence the intracavity pump power buildup is found to be of the common Airy function form:
\begin{equation}
    |A_{p,cav}/A_{p,in}|^2 = \frac{B}{1+(2F/\pi)^2\sin^2(kL/2)}
\label{eq:PumpIntracavity}
\end{equation}
where
\begin{equation}
    B = \frac{T_p}{\left(1-\sqrt{R_p (1-\ell_p)} \right)^2}
\label{eq:PumpResB}
\end{equation}
is a constant scaling factor that represents the pump power buildup on-resonance and
\begin{equation}
    \left(\frac{2F}{\pi}\right)^2 = \frac{4\sqrt{R_p (1-\ell_p)}}{\left(1-\sqrt{R_p (1-\ell_p)}\right)^2}.
\label{eq:FinesseDef}
\end{equation}
Here, $F$ represents the cavity finesse. The output power $|A_{p,out}|^2$ will be directly proportional to the intracavity power. The pump resonance curve shape is entirely determined by $F$, while any collection/detection efficiencies can be incorporated into the scaling factor $B$. Since we only want to determine $F$, we fit the resonances measured in Ext. Fig.~\ref{fig:PumpRes}b to Eq. \ref{eq:PumpIntracavity} along with a constant offset factor that comes from stray light coupling into the multimode collection fiber. 

The resultant curve fits are plotted in Ext. Fig.~\ref{fig:PumpRes}b along with the fitted value of finesse $F$. The unfitted peaks do not contribute to OPO and thus represent TM modes. The fitted value of finesse varies from $2.5\text{--}3.5$ (Ext. Fig.~\ref{fig:PumpRes}c). The variation in finesse arises because the pump cavity spectrum is sensitive to small temperature variations. From the value of finesse, Eq. \ref{eq:FinesseDef} can be solved for the total pump power recirculation, $\zeta = R_p(1-\ell_p)$. For the fitted values of finesse, $\zeta$ ranges between $8\text{--}14\%$.

Given $\zeta$, we can estimate the pump power buildup inside the cavity on resonance using Eq. \ref{eq:PumpResB} and assuming $T_p = 1-R_p$. $R_p$ is only known if we assume a value of $\ell_p$. We can reasonably assume $\ell_p$ is small and similar to the round-trip loss at telecom wavelengths ($12\%$ for measured $Q_{tot} = 1.5\times10^6$). In this regime, the pump losses are dominated by the intracavity coupler, and $R_p\approx R_p(1-\ell_p)$ (Ext. Fig.~\ref{fig:PumpRes}d). With this result in Eq. \ref{eq:PumpResB}, from Ext. Fig.~\ref{fig:PumpRes}d we find that the pump power buildup on resonance ranges from $1.75\text{--}2.15$ for the fitted values of total power recirculation $\zeta = 8\text{-}14\%$.

\subsection{Power sweep modeling}
\label{sec:PowModel}
To model the power out vs. power in data presented in Fig.~\ref{fig:fig2}b,c, we solve numerically Eq. \ref{eq:CWE} for field evolution through the periodically poled gain section for many round trips until the device reaches steady state. To implement this, for the first round trip we initialize pump, signal, and idler amplitudes inside the cavity at $z=0$ (beginning of the poled region) as:
\begin{equation}
    A^{(k=1)}(z=0) \equiv 
    \begin{pmatrix}
        A^{(1)}_p(z=0) \\
        A^{(1)}_s(z=0) \\
        A^{(1)}_i(z=0)
    \end{pmatrix}
    =
    \begin{pmatrix}
        \sqrt{P_{p,in}(1-R_p)} \\
        A_{s,0} \\
        0
    \end{pmatrix}
\end{equation}
where the superscript $k=1$ denotes the first round-trip, $A_{s,0}$ is a small value representing the random subthreshold signal field fluctuations, and the factor of $(1-R_p)$ within $A^{(1)}_p(0)$ comes because the pump power injected onto the chip $P_{p,in}$ needs to be multiplied by $(1-R_p)$ to represent the pump power that enters the periodically poled section (see Ext. Fig.~\ref{fig:PumpRes}a). Next, the three waves are propagated through the periodically poled region by numerically solving Eq. \ref{eq:CWE}, yielding $A^{(k=1)}(z=L_{pol})$, where $\eta_{DFG}$ is assumed to be $40 \text{ }\%/(\textrm{W}\cdot\textrm{cm}^2)$ (see Sec. \ref{sec:DetermineEta}) and $L_{pol} = 0.93 \textrm{ cm}$. For successive round-trips, $k>1$ and the inital amplitudes at $z=0$ are:
\begin{multline}
    A^{(k)}(0) = \\
    \begin{pmatrix}
        \sqrt{P_{p,in}(1-R_p)} + A^{(kk-1)}_p(L_{pol}) \sqrt{1-\ell_p} \sqrt{R_p} \\
        A^{(kk-1)}_s(L_{pol}) \sqrt{1-\xi_s} \\
        0
    \end{pmatrix}
\end{multline}
where $(1-\ell_p)R_p$ is the round-trip pump power recirculation, chosen to be $11\%$ (see Sec. \ref{sec:PumpResChar}), and $\xi_s \approx 12\%$ is the round-trip signal power loss based on measured $Q_{tot}$ (Fig.~1c). The idler is explicitly assumed not to resonate. We run the simulation for $N_{RT} = 5000$ round trips, which allows the system to reach steady-state. The outputs of the simulation used for Fig.~\ref{fig:fig2}b,c are steady-state idler output power $|A^{(N_{RT})}_i(L_{pol})|^2$ and steady-state intracavity signal power $|A^{(N_{RT})}_s(L_{pol})|^2$.

\subsection{Coarse tuning measurements}
The obtain the OPO coarse tuning data presented in Fig.~\ref{fig:fig3}a-c and Ext. Fig.~\ref{fig:NearDegen}e, a portion of the output light is coupled into an OSA as shown in Ext. Fig.~\ref{fig:MeasSetup}. For each temperature, the pump wavelength is tuned coarsely in $2\text{--}5$~nm steps from $1040\text{--}1090$~nm. At each coarse wavelength step, the pump wavelength is finely tuned until oscillation occurs, then wide-spanning OSA scans are taken. The noise floor of the scans is around $-70$~dBm. The peaks of the OSA scan are then extracted and plotted as in Fig.~\ref{fig:fig3}a. For temperatures above $100$~\textdegree C, the device has clean, non-degenerate output at $1.5$~µm and $3$~µm, with typical measured power around $-40$ to $-30$~dBm after being coupled into the OSA. 

\subsection{Near-degenerate OPO: measurement and simulation}
\label{sec:NearDegenOPO}

\begin{figure*}[t]
  \begin{center}
      \includegraphics[width=\textwidth]{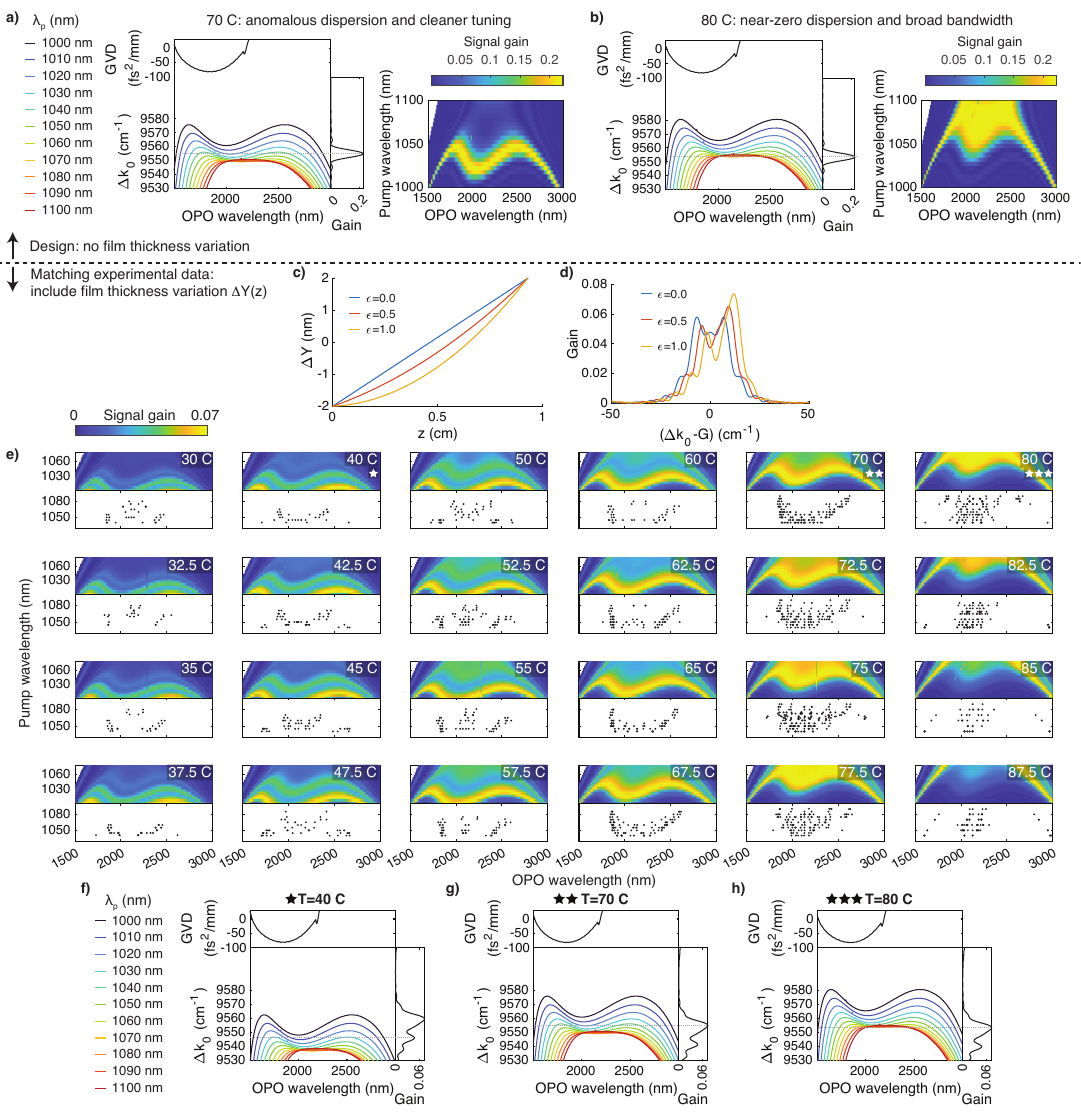}
  \end{center}
 \caption{\textbf{Near-degenerate OPO.}  \textbf{(a,b)} Dispersion-engineered OPA gain spectra near the zero-GVD point. Left panels: Simulated phase mismatch $\Delta k_0$ vs. OPO wavelength for $\lambda_p=1\text{-}1.1$~µm, alongside group-velocity dispersion (GVD) vs. OPO wavelength and OPA signal gain vs. $\Delta k_0$ (Eq. \protect\ref{eq:signal_gain_phase_mis}). Right panels: Signal gain versus $\lambda_p$ and OPO wavelength. \textbf{(a)} $70$~\textdegree C: gain curve peak intersects the region of strong anomalous dispersion, resulting in a U-shaped tuning curve. \textbf{(b)}  $80$~\textdegree C: gain curve peak intersects a region near-zero-GVD region, resulting in broadband tuning curve for $\lambda_p$ approaching $1075$~nm. \textbf{(c-h)} Verifying the dispersion engineering with experimental data.  \textbf{(c)}~Spatially-dependent, quadratic film thickness deviation $\Delta Y(z)$ for different choices of parameter $\epsilon$. \textbf{(d)} Increasing $\epsilon$ results in an asymmetric gain curve as a function of nominal phase mismatch. \textbf{(e)} Experimental OPO peaks (subpanel bottom) and simulated OPA gain spectrum (subpanel top) for $30\text{--}87.5$~\textdegree C. In both panels, y-axis is $\lambda_p$ and x-axis is OPO wavelength. \textbf{(f,g,h)} Detailed simulation plots (phase mismatch and GVD vs. OPO wavelength; OPA gain vs. phase mismatch) for \textbf{(f)} $40$~\textdegree C (side gain peak at anomalous-GVD region), \textbf{(g)} $70$~\textdegree C (main gain peak at anomalous-GVD region), and \textbf{(h)} $80$~\textdegree C (main gain peak at zero-GVD region).}
 \label{fig:NearDegen}
\end{figure*}

For temperatures below $100$~\textdegree C, the OPO approaches degenerate operation. Because of the broad gain bandwidth near degeneracy, the device can oscillate at different OPO wavelengths with slight perturbations to pump wavelength within a given coarse wavelength step. Moreover, the device can sometimes exhibit multimode oscillation with $2$ or more signal/idler mode pairs. To capture all the wavelengths the device oscillates at, we take several OSA scans for each coarse wavelength step. In Fig.~\ref{fig:fig3}a, Fig.~\ref{fig:fig3}d, and Ext. Fig.~\ref{fig:NearDegen}e we plot the locations of OSA trace peaks with peak power ${>}-45$ dBm. 

Choosing our specific waveguide geometry enables near-degenerate OPO operation near the zero-GVD point. To illustrate this, we simulate the OPO tuning curves for the design geometry: $875$~nm LN film thickness, $600$~nm etch, and $1.95$~µm top width. Simulating the modal effective index for pump wavelengths $1000\text{--}1100$~nm and signal/idler wavelengths from $1400\text{--}3500$~nm allows us to plot the phase mismatch $\Delta k_0 = k(\lambda_p) - k(\lambda_s) - k(\lambda_i)$ vs. OPO signal wavelength (Ext. Fig.~\ref{fig:NearDegen}a-b). Temperature-dependent refractive indices of LN are obtained from Umemura et. al. \cite{umemura_Sellmeier_2014, umemura_Thermooptic_2016} and of sapphire from Thomas et. al. \cite{thomas_Frequency_1998}. In both simulation temperatures $70$ and $80$~\textdegree C, the $\Delta k_0(\lambda_s)$ curves exhibit upward curvature near degeneracy for $\lambda_p\leq1060$~nm that gradually flattens as $\lambda_p$ increases to 1100~nm. The curvature of $\Delta k_0(\lambda_s)$ at degeneracy is directly related to the GVD, which can be seen by Taylor-expanding the phase mismatch around the degenerate frequency ($\equiv \Delta \omega = 0$):
\begin{align}
    \Delta k_0 (\Delta \omega) &= k(\omega_p) - k\left(\frac{\omega_p}{2} + \Delta \omega\right) - k\left(\frac{\omega_p}{2} - \Delta \omega \right) \\
    &\approx \Delta k_0(\Delta \omega=0) - \left( \beta_2 \right)_{\omega_p/2} (\Delta \omega)^2
\end{align}
where the GVD $\beta_2 = \partial^2 k /\partial\omega^2$. Hence positive curvature of $\Delta k_0$ indicates anomalous dispersion ($\beta_2 < 0$), and as $\beta_2\rightarrow 0$, the phase mismatch curves should flatten. This is verified in Ext. Fig.~\ref{fig:NearDegen}a-b by plotting GVD as a function of wavelength. 

To quasi-phasematch the process, we include the poling period $\Lambda(T)$. Incorporating the thermal expansion of LN at temperature $T\text{ } [\text{degree }C]$ and the period at $25$~\textdegree C $\Lambda_0 = 6.57$~µm, results in $\Lambda(T) = \Lambda_0[1+(1.59\times10^{-5})(T-25) + (4.9\times10^{-9})(T-25)^2]$ \cite{kim_Thermal_1969}. With the addition of the periodic poling the total phasematch becomes $\Delta k = \Delta k_0 - G$ where $G = 2\pi/\Lambda(T)$. The signal-wave gain from propagation through the periodically poled region can be found analytically by solving Eq. \ref{eq:CWE} in the presence of total phase mismatch $\Delta k$ \cite{jankowski_Dispersionengineered_2021}:
\begin{equation}
\label{eq:signal_gain_phase_mis}
    a_s(z)e^{i\Delta k z/2} = \left[ \cosh(gz) + \frac{i\Delta k}{2g}\sinh(gz)\right] a_s(0),
\end{equation} where $g = |\gamma|\sqrt{1-(\Delta k/2|\gamma|)^2}$. The simulated gain vs. $\Delta k$ results are plotted in Ext. Fig.~\ref{fig:NearDegen}a-b for $P_{pump}=600$ mW, $\eta_{DFG}=40 \%/(\textrm{W}\cdot\textrm{cm}^2)$, and $L=0.93$ cm.

Plotting the signal gain experienced for each combination of $\lambda_p$ and $\lambda_s$ constructs the OPO tuning color plots. At $70$~\textdegree C, the $\Delta k(\lambda_s)$ curves with strong upward curvature and hence anomalous dispersion experience the highest gain, leading to a U-shaped tuning curve (Ext. Fig.~\ref{fig:NearDegen}a). In contrast, at $80$~\textdegree C, the $\Delta k(\lambda_s)$ curves with the highest gain have flat curvature and hence near-zero-GVD, leading to a T-shaped tuning curve (Ext. Fig.~\ref{fig:NearDegen}b) and ultrabroad OPA gain bandwidth.

To match the experimental tuning behavior with simulation, we include a small  variation in film thickness across the poled waveguide length. Namely, we assume the LN film thickness $Y(z) = Y_0 + \Delta Y(z)$ where the nominal film thickness is $875 \textrm{~nm}$ and the spatially-dependent film thickness change $\Delta Y(z) = -\Delta Y_{tot}/2 + bz + az^2$ where the total film thickness variation $\Delta Y_{tot} = 4\textrm{~nm}$, $b = b_0 - aL$, $a = \epsilon b_0$ and $b_0 = (\Delta Y(L) - \Delta Y(0))/L$. The $4\textrm{~nm}$ simulated total film thickness variation over $9.3$~mm length was chosen as it is the minimum film thickness where simulated results qualitatively match experiment. Moreover, the chosen thickness variation agrees well with thickness measurements performed by the LN-on-sapphire vendor, which indicate around $0.4$~nm LN thickness variation per mm length. The factor $\epsilon$ describes the curvature of the film thickness variation, as depicted in Ext. Fig.~\ref{fig:NearDegen}c. To calculate how film thickness variation profiles affect the signal wave gain, we solve Eq. \ref{eq:CWE} numerically in the presence of spatially-dependent phase mismatch $\Delta k = \Delta k_0  - G + \Delta k_h(z)$ where the phase mismatch due to height variations $\Delta k_h(z) = \frac{d\Delta k_{h}}{dY} \Delta Y(z)$ and the ratio of phase mismatch shift to change in film thickness $\frac{d\Delta k_{h}}{dY} = 8.5 \textrm{ cm}^{-1} / \textrm{nm}$ is found from simulation. Specifically, we solve:
\begin{align}
\frac{d}{dz}
a_s 
&=
\gamma \exp\left[-i\int_0^z \Delta k(z')dz' \right] a_i^* \\
\frac{d}{dz}
a_i^*
&=
\gamma^* \exp\left[i\int_0^z \Delta k(z')dz' \right] a_s
\end{align}
The resultant signal gain as a function of the constant part of phase mismatch $\Delta k_0 - G$ is plotted in Ext. Fig.~\ref{fig:NearDegen}d. For linear film thickness variation ($\epsilon = 0$), the gain curve vs. phase mismatch has broadened, reduced in magnitude, and exhibits two major peaks instead of one major peak found when $\Delta k_h(z) = 0$ (Ext. Fig.~\ref{fig:NearDegen}a,b). The addition of quadratic film thickness variation ($\epsilon > 0$) makes the gain curve slightly asymmetric, which matches experimental results. 

Simulated OPO tuning curves along with experimental data for $T = 30\text{--}87.5$~\textdegree C, $\Lambda_0 = 6.58$~µm, and $\epsilon=1$ are shown in Ext. Fig.~\ref{fig:NearDegen}e. The experimental data qualitatively matches simulation. As temperature increases in both simulation and experiment, the observed OPO output tuning curves shift upwards in the plot (towards longer pump wavelengths). $40$~\textdegree C (labeled with $\star$) and $70$~\textdegree C ($\star \star$) both present U-shaped tuning curves, while $80$~\textdegree C ($\star \star \star$) presents a T-shaped tuning curve. To understand the results, we highlight the simulation results of $40$~\textdegree C, $70$~\textdegree C, and $80$~\textdegree C (Ext. Fig.~\ref{fig:NearDegen}f-h). At $40$~\textdegree C, the U-shaped tuning curve arises from a secondary gain peak amplifying anomalous dispersion regions (Ext. Fig.~\ref{fig:NearDegen}f). By $70$~\textdegree C, a stronger U-shaped tuning curve arises from the main gain peak amplifying the same anomalous dispersion regions (Ext. Fig.~\ref{fig:NearDegen}g). Finally, at $80$~\textdegree C, the U-shape transforms into a T-shape when the main gain peak amplifies regions of near-zero-GVD, leading to broadband OPO in both simulation and experiment (Ext. Fig.~\ref{fig:NearDegen}h).

\subsection{Fine tuning characterization}

\begin{figure*}[t]
  \begin{center}
      \includegraphics[width=\textwidth]{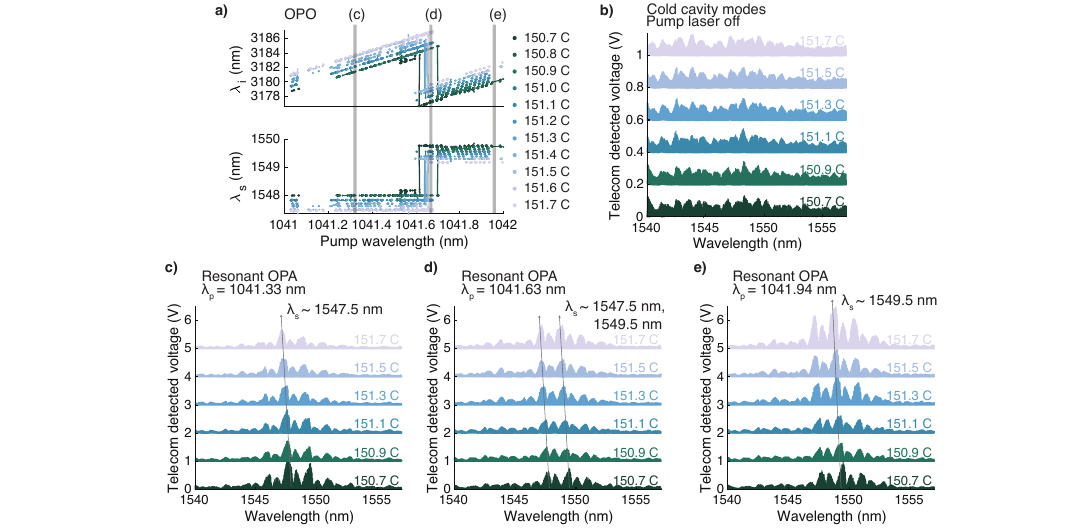}
  \end{center}
 \caption{\textbf{Understanding the OPO fine tuning.} \textbf{(a)} Measured OPO signal and idler wavelengths as pump wavelength is tuned over $1$~nm, for different temperatures from $150.7\text{-}151.7$~\textdegree C. \textbf{(b)} Cold cavity (pump laser off) telecom mode spectra for temperatures from $150.7\text{-}151.7$~\textdegree C. \textbf{(c, d, e)} Resonant OPA scans at a fixed pump wavelength, which illustrate the net gain that each of the telecom-wavelength cavity modes experience for different temperatures and pump wavelength. \textbf{(c)} At pump wavelength $1041.33$~nm, the modes near $1547.5$~nm are preferred. \textbf{(d)} For pump wavelength $1041.63$~nm, modes near $1547.5$ and $1549.5$~nm both experience high gain, explaining the device’s large mode hop. \textbf{(e)} For pump wavelength $1041.94$~nm, modes near $1549.5$~nm are preferred. }
 \label{fig:FineTun}
\end{figure*}

OPO fine tuning data (Fig.~\ref{fig:fig3}e, Ext. Fig.~\ref{fig:FineTun}a) is obtained by taking ${\sim}10$ $100$-pm-wide pump wavelength piezo scans and recording the wavelength of both pump and generated OPO signal. To do so, the measurement setup in Ext. Fig.~\ref{fig:MeasSetup} is modified; generated telecom-wavelength OPO output is collected in a single-mode lensed fiber to increase wavelength resolution. Directly connecting this fiber to a rapidly-scanning OSA allows determination of the telecom-band wavelength. The plotted idler wavelength is calculated based on energy conservation. OPO wavelengths collected during a single OPO ``cluster" (Fig.~\ref{fig:fig2}a) are plotted as dots connected by lines. Data is recorded at $11$ temperatures between $150.7\text{--}151.7$~\textdegree C. 

The presented fine tuning data (Ext. Fig.~\ref{fig:FineTun}a) exhibits three regimes: (1) Clean tuning from $\lambda_p=1041\text{--}1041.5$~nm at $\lambda_s\approx1547.5$~nm, (2) Strong mode hop at $\lambda_p=$ $1041.6$~nm between $\lambda_s\approx 1547.5, 1549.5$~nm, and (3) Clean tuning from $\lambda_p=1041.7\text{--}1042$~nm at $\lambda_s\approx1549.5$~nm. The signal mode transition from $\lambda_s = 1547.5$~nm $\rightarrow 1549.5$~nm as $\lambda_p = 1041$~nm $\rightarrow 1042$~nm comes because changing $\lambda_p$ shifts the nonlinear gain spectrum.

To clarify how the shift in gain spectrum affects which modes oscillate, we measure the amplification experienced by cavity modes (Ext. Fig.~\ref{fig:FineTun}b-e). With no pump laser power, the cavity mode spectrum is obtained by sweeping the wavelength of the tunable telecom laser (Santec TSL-710) input to the chip and detecting the output in an InGaAs photodiode (Newport 1623). The reduced-height peaks in the mode-spectrum structure comes from mode crossings. Then we pump the device below the OPO threshold while scanning the tunable telecom laser and measuring the output at telecom wavelengths. The pump laser amplifies all the telecom cavity modes within its gain bandwidth (Ext. Fig.~\ref{fig:FineTun}c-e). The highest-peaked modes in these plots are those that experienced the highest net gain and thus oscillate when the device is above threshold. When $\lambda_p=1041.33$~nm (Ext. Fig.~\ref{fig:FineTun}c), the cavity modes near $1547.5$~nm experience the most gain. When $\lambda_p$ shifts near $1041.6$~nm, two groups of modes near $1547.5$ and $1549.5$~nm compete for gain (Ext. Fig.~\ref{fig:FineTun}d), explaining the mode crossing observed in (Ext. Fig.~\ref{fig:FineTun}a). Finally, when $\lambda_p= 1041.94$~nm, the modes at $1549.5$~nm have high gain. 

\subsection{Methane OPO spectroscopy}

As shown in Fig.~\ref{fig:fig4}a, the measurement setup for methane spectroscopy slightly modifies the general setup shown in Ext. Fig.~\ref{fig:MeasSetup}. The methane cell ($7.5$~cm length, $20$~Torr, Triad Technologies) is placed before a MCT detector (Thorlabs PDAVJ5), while the reference arm uses an InAsSb detector (Thorlabs PDA07P2). The generated signal wavelength is measured by heterodyning it with blueshifted output from a tunable telecom laser (Santec TSL-710) on a $12$~GHz fast photodiode (Newport 1554-B). The wavelength of the reference telecom laser is read directly before and after a given measurement sweep using the wavemeter (Bristol Instruments Model 621). The beatnote is read using an RF spectrum analyzer (Rohde and Schwartz FSW26). 

In Fig.~\ref{fig:fig4}c we present methane absorption features for four separate scans. The MIR wavelength axis is obtained by combining the interpolated $\lambda_p$ and $\lambda_s$ axes by energy conservation. The presented scans in Fig.~4c are background-corrected to match the theoretical curves from HITRAN by fitting the experimental transmission data to $T(\lambda) = V_1(\lambda) / V_2(\lambda) = x_1 e^{-\alpha(\lambda) L} + x_2$ where the fitting parameter $x_1$ accounts for transmission/detection efficiency differences in the two paths generating voltages $V_1(t)$ and $V_2(t)$, while $x_2$ accounts for MIR light hitting detector $V_1$ that did not couple into the gas cell (the MIR beam diameter was $\sim$2x the gas cell diameter when viewed with an IR viewing card). The background fitting process did not affect the wavelength axis. As a result, the experimental backround-subtracted data plotted in Fig.~\ref{fig:fig4}c is $(T(\lambda) - x_2)/x_1$. HITRAN data plotted in Fig.~\ref{fig:fig4}c comes from calculating absorption coefficient $\alpha(\lambda)$ from the HITRAN database for methane at $20$~Torr, then plotting $T=\exp(-\alpha(\lambda) L)$. 

\subsection{Resonant DFG spectroscopy}

For resonant DFG spectroscopy, we use the same setup as for methane OPO spectroscopy but instead use a gas cell of ammonia ($1.5$~cm length, $740$~Torr, purchased from Wavelength References). $\lambda_p$ is determined with a wavemeter before scanning $\lambda_s$ from $1540\text{--}1620$~nm, which is assumed to vary linearly across the scan range. Generated experimental data, consisting of discrete peaks of generated MIR output (see Fig.~\ref{fig:fig4}e), is processed by fitting Lorentzians to each peak, then calculating the transmission for the $k$-th peak $T(\lambda_k)$ based on the ratio of peak areas from the sample path and reference path. ${\sim}10$ scans are taken, then averaged, to improve SNR. Experimental absorbance data plotted in Fig.~\ref{fig:fig4}f is $A=-\log(T(\lambda_k)/T_{bg})$ where $T_{bg} = 0.7$ to account for the difference in transmission between sample and reference paths. HITRAN data plotted in Fig.~\ref{fig:fig4}f is obtained by calculating the absorption coefficient $\alpha(\lambda)$ from the HITRAN database for ammonia at $740$~Torr then plotting $A = \alpha(\lambda) L$.

\section*{Acknowledgements} 
 We thank NTT Research for their financial and technical support. We thank the United States government for their support through the Department of Energy Grant No. DE-AC02-76SF00515, the Defense Advanced Research Projects Agency (DARPA) LUMOS program (Grant No. HR0011-20-2-0046), the DARPA Young Faculty Award (YFA, Grant No. D19AP00040), the U.S. Department of Energy (Grant No. DE-AC02-76SF00515) and Q-NEXT NQI Center, and the U.S. Air Force Office of Scientific Research MURI grant (Grant No. FA9550-17-1-0002). A.Y.H. acknowledges NSF GRFP, Grant. No. 2146755. H.S.S. acknowledges support from the Urbanek Family Fellowship, and V.A. was partially supported by the Stanford Q-Farm Bloch Fellowship Program and the Max Planck Institute in Erlangen. This work was also performed at the Stanford Nano Shared Facilities (SNSF), supported by the National Science Foundation under award ECCS-2026822. We also acknowledge the Q-NEXT DOE NQI Center and the David and Lucille Packard Fellowship for their support. We thank Leo Hollberg for many useful discussions and lending the cells for the gas spectroscopy experiment. 

\section*{Author contributions}
A.Y.H., H.S., C.L., V.A., and A.H.S-N. designed the device. 
A.Y.H., H.S, and T.P. fabricated the device.
A.Y.H, H.S., T.P.M., and T.P. developed fabrication procedures together. 
A.Y.H. measured the device. A.Y.H. analyzed data with support from H.S., M.J., and J.M.
M.M.F. and A.H.S.-N. advised the project and provided experimental/theoretical support.
A.Y.H. drafted the manuscript with input from all the authors.

\end{document}